\documentclass[preprint,showpacs,preprintnumbers,amsmath,amssymb]{revtex4}
\usepackage{graphicx}
\usepackage{subfig}
\usepackage{slashed}
\usepackage{color}
\definecolor{red}{rgb}{1,0,0}

\newcommand{\tvect}[2]{%
  \ensuremath{\Bigl(\negthinspace\begin{smallmatrix}#1\\#2\end{smallmatrix}\Bigr)}}

\begin{document}

\title{Majorana neutrino decay in an Effective Approach}

\author{Luc\'{\i}a Duarte}
\email{lduarte@fing.edu.uy}
 \affiliation{Instituto de F\'{\i}sica, Facultad de Ingenier\'{\i}a,
 Universidad de la Rep\'ublica \\ Julio Herrera y Reissig 565,(11300) 
Montevideo, Uruguay.}

\author{Javier Peressutti}
\affiliation{Instituto de F\'{\i}sica de Mar del Plata (IFIMAR)\\
CONICET, UNMDP\\ Departamento de F\'{\i}sica,
Universidad Nacional de Mar del Plata \\
Funes 3350, (7600) Mar del Plata, Argentina}
 
\author{Oscar A. Sampayo}
\email{sampayo@mdp.edu.ar}

 \affiliation{Instituto de F\'{\i}sica de Mar del Plata (IFIMAR)\\ CONICET, UNMDP\\ Departamento de F\'{\i}sica,
Universidad Nacional de Mar del Plata \\
Funes 3350, (7600) Mar del Plata, Argentina}

\begin{abstract}
The search strategy or the discovery of new effects for heavy neutrinos often rely on their different decay channels to detectable particles. In particular in this work we study the decay of a Majorana neutrino with interactions obtained from an effective general theory modeling new physics at the scale $\Lambda$. The results obtained are general because they are based in an effective theory and not in specific models. We are interested in relatively light heavy Majorana neutrinos, with masses lower than the $W$ mass ($m_N<m_W$). This mass range simplifies the study by reducing the possible decay modes. Moreover, we found that for $\Lambda\sim 1$ T$e$V, the neutrino plus photon channel could account for different observations: we analyze the potentiality of the studied interactions to explain some neutrino-related problems like the MiniBooNE and SHALON anomalies. We show in different figures the dominant branching ratios and the decay length of the Majorana neutrino in this approach. This kind of heavy neutral leptons could be searched for in the LHC with the use of displaced vertices techniques. \
\end{abstract}

\maketitle

\section{\bf Introduction}

One of the most spectacular new results in high energy physics is the discovery of neutrino oscillations, indicating that they are not massless. 
The neutrinos can be of two different types: either Dirac or Majorana particles. Dirac fermions have distinct particle and antiparticle degrees of freedom while Majorana fermions make no such distinction and have half as many degrees of freedom \cite{Kayser:1989iu}. In this conditions fermions with conserved charges (color, electric charge, lepton number,...) must be of Dirac type, while fermions without conserved charges may be of either type. New  still undetected neutrinos could have large masses and be of either type. If  heavy neutrinos ($N$) do exist, present and future experiments would offer the possibility of establishing their nature. The production of Majorana neutrinos via $e^+e^-$, $e^- \gamma$, $\gamma \gamma$, $e^- P$ and hadronic collisions have been extensively investigated \cite{Ma:1989jpa, Datta:1993nm, Gluza:1994ac, Hofer:1996cs, Cvetic:1998vg, Almeida:2000pz, Peressutti:2001ms, Peressutti:2002nf, Peressutti:2011kx, Peressutti:2014lka, Atre:2009rg,  Buchmuller:1991tu, Blaksley:2011ey, Duarte:2014zea, Antusch:2015mia, Deppisch:2015qwa, delAguila:2006dx, delAguila:2008ir}.

A very known scenario for the study of Majorana neutrinos is the seesaw mechanism \cite{Mohapatra:1998rq}, requiring the existence of at least one type of heavy right-handed Majorana neutrino. As indicated in \cite{delAguila:2008ir}, the parameters determining the interaction of the heavy Majorana neutrino $N$ with the standard particles turn out to be very small, indicating the need for a new approach involving physics beyond the typical seesaw scenarios. 

In this work we study the decay modes of a relatively light heavy Majorana neutrino in the context of a general effective framework. We focus in a mass interval below the standard massive vector bosons mass ($m_{N} < m_{W}$) as this reduces the possible decay channels, letting us concentrate on the phenomenology of the neutrino plus photon mode. This heavy neutrino decay channel has  been introduced as a possible answer to some experimental puzzles, like the MiniBooNE \cite{AguilarArevalo:2007it, AguilarArevalo:2008rc} and SHALON \cite{Sinitsyna:2013hmn} anomalies, considering sterile heavy neutrinos created by $\nu_{\mu}$ neutral current interactions and decaying radiatively due a transition magnetic moment \cite{Gninenko:2009ks}. We revisit here the mentioned anomalies in the light of an effective Lagrangian description for the heavy Majorana neutrino decays.  

The paper is organized as follows: in subsections \ref{EffL}, \ref{Decay_Widths} and \ref{Bounds} we describe the model-independent effective approach, and show our analytic and numerical results for the Majorana neutrino decay widths and branching ratios, and discuss the existing bounds on the effective couplings. In section \ref{applications} we explore the potentiality of the effective approach to explain the MiniBooNE and SHALON anomalies, showing our results for the Majorana neutrino lifetime and decay length. We present our conclusions in section \ref{conclusions}.  

\subsection{Effective Lagrangian}\label{EffL}

As it was explained in \cite{delAguila:2008ir}, the presence of Majorana neutrinos would be a signal of physics beyond the minimal seesaw mechanism, and thus their interactions would be best described in a model-independent effective approach. 
\begin{figure*}[htbp]
\centering
\includegraphics[totalheight=8cm]{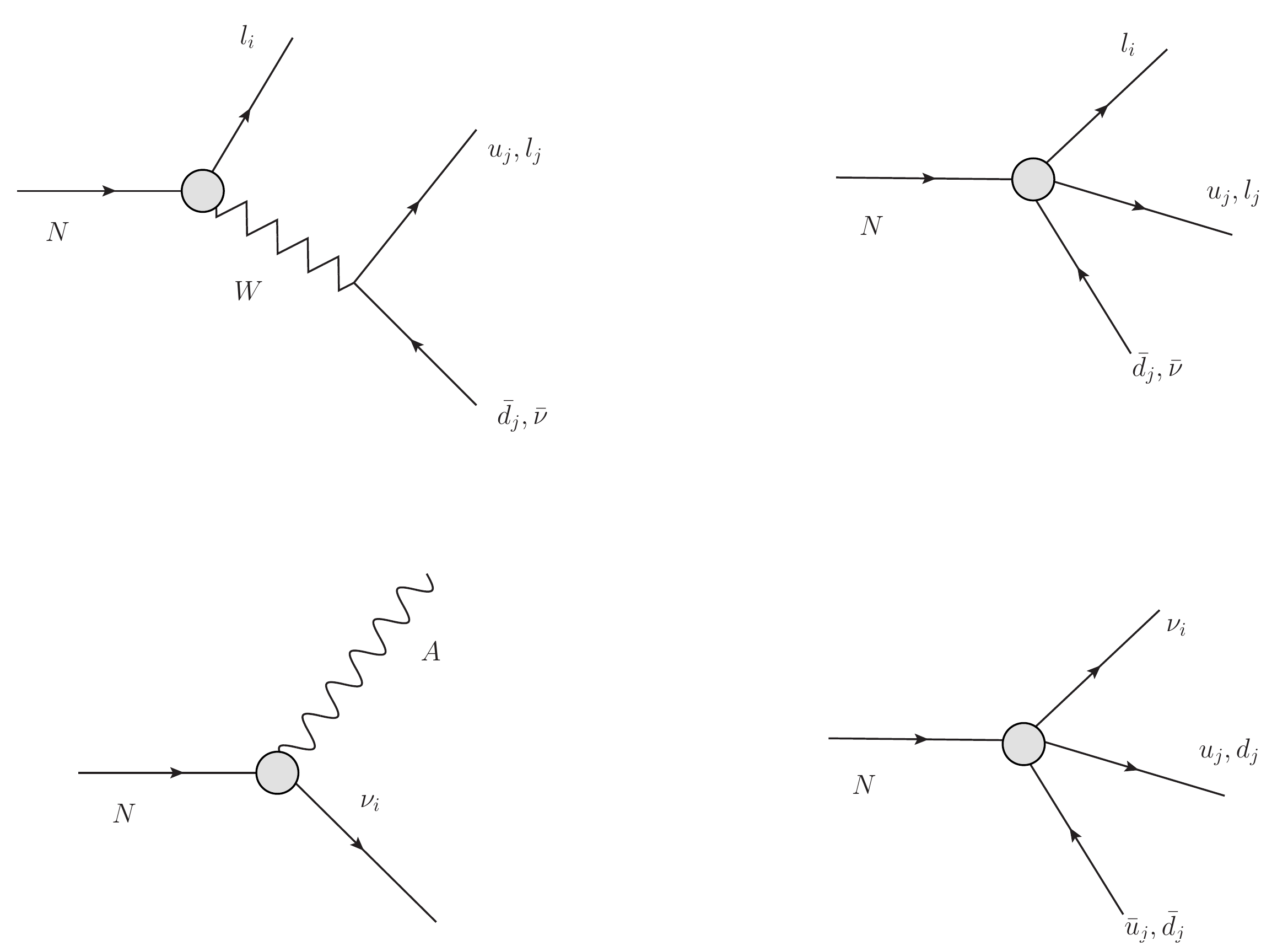}
\caption{\label{fig:Ndecay} Decay modes for Majorana neutrinos with $m_N<m_W$.}
\end{figure*}
It is possible to parametrize the effects of new physics beyond the standard model by a  set of effective operators $\mathcal{O}$ constructed with the standard model and the Majorana neutrino fields and satisfying the Standard Model $SU(2)_L \otimes U(1)_Y$ gauge symmetry \cite{delAguila:2008ir}.  The effect of these operators is suppressed by inverse powers of the new physics scale $\Lambda$, for which we take the value $\Lambda = 1$ T$e$V. The total Lagrangian is organized as follows:
\begin{eqnarray}
\mathcal{L}=\mathcal{L}_{SM}+\sum_{n=6}^{\infty}\frac1{\Lambda^{n-4}}\sum_i
\alpha_i \mathcal{O}_i^{(n)}
\end{eqnarray}
For the considered operators we follow \cite{delAguila:2008ir} starting with a rather general effective Lagrangian density for the interaction of right handed Majorana neutrinos $N$ with bosons, leptons and quarks. The following are dimension $6$ operators and could be generated at tree level in the unknown fundamental ultraviolet theory. 
The first subset includes operators with scalar and vector bosons (SVB),
\begin{eqnarray} \label{eq:ope1}
\mathcal{O}_{LN\phi}=(\phi^{\dag}\phi)(\bar L_i N \tilde{\phi}),
\;\; \mathcal{O}_{NN\phi}=i(\phi^{\dag}D_{\mu}\phi)(\bar N \gamma^{\mu} N), \;\; \mathcal{O}_{Ne\phi}=i(\phi^T \epsilon D_{\mu}
\phi)(\bar N \gamma^{\mu} l_i)
\end{eqnarray}
and a second subset includes the baryon-number conserving 4-fermion contact terms:
\begin{eqnarray} \label{eq:ope2}
\mathcal{O}_{duNe}=(\bar d_i \gamma^{\mu} u_i)(\bar N \gamma_{\mu}
l_i) &,& \;\; \mathcal{O}_{fNN}=(\bar f_i \gamma^{\mu} f_i)(\bar N
\gamma_{\mu}
N), \\
\mathcal{O}_{LNLe}=(\bar L_i N)\epsilon (\bar L_i l_i)&,& \;\; \nonumber
\mathcal{O}_{LNQd}=(\bar L_i N) \epsilon (\bar Q_i d_i), \cr
\mathcal{O}_{QuNL}=(\bar Q_i u_i)(\bar N L_i)&,& \;\;
\mathcal{O}_{QNLd}=(\bar Q_i N)\epsilon (\bar L_i d_i),\cr
\mathcal{O}_{LN}=|\bar N L_i|^2&&
\end{eqnarray}
where $l_i$, $u_i$, $d_i$ and $L_i$, $Q_i$ denote, for the family
labeled $i$, the right handed $SU(2)$ singlet and the left-handed
$SU(2)$ doublets, respectively.
In addition, there are operators generated at one-loop level in the underlying full theory whose coefficients are
naturally suppressed by a factor $1/16\pi^2$ \cite{delAguila:2008ir, Arzt:1994gp}:
\begin{eqnarray}
\mathcal{O}^{(5)}_{NNB} & = & \bar N \sigma^{\mu\nu} N^c B_{\mu\nu}, \cr
\mathcal{O}_{ N B} = (\bar L \sigma^{\mu\nu} N) \tilde \phi B_{\mu\nu} , &&
\mathcal{O}_{ N W } = (\bar L \sigma^{\mu\nu} \tau^I N) \tilde \phi W_{\mu\nu}^I , \cr
\mathcal{O}_{ D N} = (\bar L D_\mu N) D^\mu \tilde \phi, &&
\mathcal{O}_{ \bar D N} = (D_\mu \bar L N) D^\mu \tilde \phi \ .
\label{eq:ope3}
\end{eqnarray}
Taking the scalar doublet after spontaneous symmetry breaking as $\phi=\tvect{0}{\frac{v+h}{\sqrt{2}}}$, with $h$ being the Higgs field, the operators listed in (\ref{eq:ope1}) contribute to the effective Lagrangian  
\begin{eqnarray}\label{leff_svb}
 \mathcal{L}^{tree}_{SVB} &= & \frac{1}{\Lambda^2}\left\{\alpha^{(i)}_{\phi}
\left( \frac{3v^2}{2\sqrt{2}}~\bar \nu_{L,i} N_R~ h + \frac{3v}{2\sqrt{2}}~\bar \nu_{L,i} N_R~ h h+ \frac{1}{2\sqrt{2}}~\bar \nu_{L,i} N_R~ h h h \right) \right. 
\nonumber
\\ && \left. - \alpha_Z \left( -(\bar N_R \gamma^{\mu} N_R) \left( \frac{m_Z}{v} Z_{\mu} \right) \left( \frac{v^2}{2} + v h + \frac{1}{2} h h \right) \right. \right.
\nonumber 
\\ &&  \left. \left. + (\bar N_R \gamma^{\mu} N_R) \left( \frac{v}{2} P^{(h)}_{\mu} h +\frac{1}{2} P^{(h)}_{\mu} h h \right)  \right) \right.
\nonumber
\\ && - \left. \alpha^{(i)}_W (\bar N_R \gamma^{\mu} l_R)\left(\frac{v m_{W}}{\sqrt{2}}W^{+}_{\mu} + \sqrt{2} m_{W} W^{+}_{\mu} h + \frac{g}{2 \sqrt{2}} W^{+}_{\mu} h h \right) + h.c. \right\}.
\end{eqnarray}
The 4-fermion Lagrangian can be written \eqref{eq:ope2}:
\begin{eqnarray}\label{leff_4-f}
\mathcal{L}^{tree}_{4-f}&=& \frac{1}{\Lambda^2} \left\{ \alpha^{(i)}_{V_0} \bar d_{R,i} \gamma^{\mu} u_{R,i} \bar N_R \gamma_{\mu}
l_{R,i} + \alpha^{(i)}_{V_1} \bar l_{R,i} \gamma^{\mu} l_{R,i} \bar
N_R \gamma_{\mu} N_R + \alpha^{(i)}_{V_2} \bar L_i \gamma^{\mu} L_i
\bar N_R \gamma_{\mu} N_R + \right. \nonumber
\\ && \left. \alpha^{(i)}_{V_3} \bar u_{R,i} \gamma^{\mu}
u_{R,i} \bar N_R \gamma_{\mu} N_R + \alpha^{(i)}_{V_4} \bar d_{R,i}
\gamma^{\mu} d_{R,i} \bar N_R \gamma_{\mu} N_R + \alpha^{(i)}_{V_5}
\bar Q_i \gamma^{\mu} Q_i \bar N_R \gamma_{\mu} N_R + \right.
\nonumber
\\ && \left.
\alpha^{(i)}_{S_0}(\bar \nu_{L,i}N_R \bar e_{L,i}l_{R,i}-\bar
e_{L,i}N_R \bar \nu_{L,i}l_{R,i}) + \alpha^{(i)}_{S_1}(\bar
u_{L,i}u_{R,i}\bar N \nu_{L,i}+\bar d_{L,i}u_{R,i} \bar N e_{L,i})
 + \right. \nonumber
\\ && \left.
\alpha^{(i)}_{S_2} (\bar \nu_{L,i}N_R \bar d_{L,i}d_{R,i}-\bar
e_{L,i}N_R \bar u_{L,i}d_{R,i}) + \alpha^{(i)}_{S_3}(\bar u_{L,i}N_R
\bar e_{L,i}d_{R,i}-\bar d_{L,i}N_R \bar \nu_{L,i}d_{R,i}) + \right.
\nonumber
\\ && \left.  \alpha^{(i)}_{S_4} (\bar N_R \nu_{L,i}~\bar l_{L,i} N_R~+\bar
N_R e_{L,i} \bar e_{L,i} N_R)  + h.c. \right\}
\end{eqnarray}
In Eqs. \eqref{leff_svb} and \eqref{leff_4-f} a sum over the family index $i$ is understood, and the
constants $\alpha^{(i)}_{\mathcal O}$ are associated to specific operators:
\begin{eqnarray}
\alpha_Z&=&\alpha_{NN\phi},\;
\alpha^{(i)}_{\phi}=\alpha^{(i)}_{LN\phi},\;
\alpha^{(i)}_W=\alpha^{(i)}_{Ne\phi},\;
\alpha^{(i)}_{V_0}=\alpha^{(i)}_{duNe},\;\;
\alpha^{(i)}_{V_1}=\alpha^{(i)}_{eNN},\;\nonumber \\
\alpha^{(i)}_{V_2}&=&\alpha^{(i)}_{LNN},\;\alpha^{(i)}_{V_3}=\alpha^{(i)}_{uNN},\;
\alpha^{(i)}_{V_4}=\alpha^{(i)}_{dNN},\;\alpha^{(i)}_{V_5}=\alpha^{(i)}_{QNN},\;
\alpha^{(i)}_{S_0}=\alpha^{(i)}_{LNe},\;\nonumber \\
\alpha^{(i)}_{S_1}&=&\alpha^{(i)}_{QuNL},\;
\alpha^{(i)}_{S_2}=\alpha^{(i)}_{LNQd},\;\;
\alpha^{(i)}_{S_3}=\alpha^{(i)}_{QNLd},\;
\alpha^{(i)}_{S_4}=\alpha^{(i)}_{LN}.
\end{eqnarray}

For the case of the one-loop generated operators in \eqref{eq:ope3}, we have the effective Lagrangian:

\begin{eqnarray}\label{leff_1loop}
\mathcal{L}_{eff}^{1-loop}&=&\frac{\alpha_{L_1}^{(i)}}{\Lambda^2} \left(-i\sqrt{2} v c_W P^{(A)}_{\mu} ~\bar \nu_{L,i} \sigma^{\mu\nu} N_R~ A_{\nu} 
+i \sqrt{2} v s_W P^{(Z)}_{\mu} ~\bar \nu_{L,i} \sigma^{\mu\nu} N_R~ Z_{\nu}+  \right.
\nonumber 
\\ && \left. -i\sqrt{2} c_W P^{(A)}_{\mu} ~\bar\nu_{L,i} \sigma^{\mu\nu} N_R~ A_{\nu} h + i \sqrt{2}  s_W P^{(Z)}_{\mu} ~\bar \nu_{L,i} \sigma^{\mu\nu} N_R~ Z_{\nu} h \right)  
\nonumber 
\\ && -\frac{\alpha_{L_2}^{(i)}}{\Lambda^2} \left(\frac{m_Z}{\sqrt{2}}P^{(N)}_{\mu} ~\bar \nu_{L,i} N_R~ Z^{\mu}+ \frac{m_{z}}{\sqrt{2}v} P^{(N)}_{\mu} ~\bar \nu_{L,i} N_R~ Z^{\mu} h  \right.
\nonumber
\\ &&  \left. + m_W P^{(N)}_{\mu} ~\bar l_{L,i} N_R~ W^{-\mu} + \frac{\sqrt{2} m_{W}}{v} P^{(N)}_{\mu} ~\bar l_{L,i} N_R~ W^{-\mu} h + \frac{1}{\sqrt{2}} P^{(h)}_{\mu} P^{(N)\mu} ~\bar \nu_{L,i} N_R~ h \right) 
\nonumber 
\\ && -\frac{\alpha_{L_3}^{(i)}}{\Lambda^2}\left(i\sqrt{2} v  c_W P^{(Z)}_{\mu} ~\bar \nu_{L,i} \sigma^{\mu\nu}N_R~ Z_{\nu} 
+ i\sqrt{2} v s_W P^{(A)}_{\mu} ~\bar \nu_{L,i} \sigma^{\mu\nu}N_R~ A_{\nu} \right.
\nonumber 
\\ && \left. +i 2\sqrt{2} m_W ~\bar \nu_{L,i} \sigma^{\mu\nu} N_R~ W^+_{\mu}W^-_{\nu} + i \sqrt{2} v P^{(W)}_{\mu} ~\bar l_{L,i} \sigma^{\mu\nu} N_R~ W^-_{\nu} \right.
\nonumber
\\ && \left. + i 4 m_W c_W ~\bar l_{L,i} \sigma^{\mu\nu} N_R~ W^-_{\mu} Z_{\nu}+ i 4 m_W s_W ~\bar l_{L,i} \sigma^{\mu\nu} N_R~ W^-_{\mu} A_{\nu} \right.
\nonumber 
\\ && \left. + i \sqrt{2} P^{(W)}_{\mu} ~\bar l_{L,i} \sigma^{\mu\nu} N_R~ W^-_{\nu} h + i 2 g  c_W ~\bar l_{L,i} \sigma^{\mu\nu} N_R~ W^-_{\nu} Z_{\mu} h \right.
\nonumber 
\\ && \left. + i 2 g s_W ~\bar l_{L,i} \sigma^{\mu\nu} N_R~ W^-_{\nu} A_{\mu} h  + i \sqrt{2} c_W P^{(Z)}_{\mu} ~\bar \nu_{L,i} \sigma^{\mu\nu} N_R~ Z_{\mu} h \right.
\nonumber
\\ && \left. + i \sqrt{2} s_W P^{(A)}_{\mu} ~\bar \nu_{L,i} \sigma^{\mu\nu} N_R~ A_{\mu} h +i \sqrt{2} g ~\bar \nu_{L,i} \sigma^{\mu\nu} N_R~ W^{+}_{\mu} W^{-}_{\nu} h
\right) 
\nonumber
\\ &&  -\frac{\alpha_{L_4}^{(i)}}{\Lambda^2} \left( \frac{m_Z}{\sqrt{2}} P^{(\bar\nu)}_{\mu}~\bar \nu_{L,i} N_R~ Z_{\mu}+ \frac{m_{Z}}{\sqrt{2}v} (P^{(\bar\nu)}_{\mu}-P^{(h)}_{\mu})~\bar \nu_{L,i} N_R~ Z^{\mu} h  \right. 
\nonumber
\\ && \left. + \frac{1}{\sqrt{2}} P^{(h) \mu} P^{(\bar\nu)}_{\mu}~\bar \nu_{L,i} N_R~ h -\frac{\sqrt{2} m^2_W}{v} ~\bar \nu_{L,i} N_R~  W^{-\mu}W^+_{\mu} 
-\frac{m^2_z}{\sqrt{2} v} ~\bar \nu_{L,i} N_R~ Z_{\mu}Z^{\mu}   \right.
\nonumber
\\ && \left. -\frac{1}{2}\frac{m^2_{Z}}{v^2} ~\bar \nu_{L,i} N_R~ Z_{\mu}Z^{\mu} h -\frac{\sqrt{2} m^2_W}{v^2} ~\bar \nu_{L,i} N_R~ W^{+}_{\mu} W^{-\mu} h \right.
\nonumber
\\ && \left. + m_W P^{(\bar l)}_{\mu} W^{-\mu} ~\bar l_{L,i} N_R~ +\frac{m_W}{v} (P^{(\bar l)}_{\mu}-P^{(h)}_{\mu}) W^{-\mu} ~\bar l_{L,i} N_R~ h \right.
\nonumber
\\ && \left. + e m_W   ~\bar l_{L,i} N_R~ W^{-\mu}A_{\mu}  + e m_Z s_W ~\bar l_{L,i} N_R~ W^{-\mu}Z_{\mu}  \right.
\nonumber
\\ &&  \left. + \frac{e m_Z s_W}{v} ~\bar l_{L,i} N_R~ Z_{\mu} W^{-\mu} h + \frac{e m_Z c_W}{\sqrt{2} v} ~\bar l_{L,i} N_R~ A_{\mu} W^{-\mu} h\right) + h.c.
\end{eqnarray}
where $P^{(a)}$ is the 4-moment of the incoming $a$-particle and a sum over the family index $i$ is understood again. 
The constants $\alpha^{(i)}_{L_j}$ with $j=1,2,3,4$ are associated to the specific operators:
\begin{eqnarray}
\alpha^{(i)}_{L_1}=\alpha^{(i)}_{NB},\;\; \alpha^{(i)}_{L_2}=\alpha^{(i)}_{DN},\;\; \alpha^{(i)}_{L_3}=\alpha^{(i)}_{NW},\;\; 
\alpha^{(i)}_{L_4}=\alpha^{(i)}_{\bar DN}.
\end{eqnarray}

\subsection{Decay Widths}\label{Decay_Widths}

We have calculated the decay channels for a Majorana neutrino with mass lower than the standard model vector bosons $(m_{N}<m_{W})$. 
This range allows for the decay to fermions (excepting the top quark) and to photons. The contributing decay modes are schematically shown in Fig.\ref{fig:Ndecay}.

We now present the partial decay widths of a heavy Majorana neutrino $N$ decaying to three fermions. They were calculated using the effective Lagrangian \eqref{leff_4-f}.

The decays to one lepton and two quarks can be written:
\begin{eqnarray}\label{dquarks_1}
\frac{d\Gamma}{dx}^{(N\rightarrow l^{+} \bar u d)}&=&\frac{m_N }{512
\pi^3}\left(\frac{m_N}{\Lambda}\right)^4  x \frac{(1-x-y_l+y_u)}{(1-x+y_l)^ 3}
 \left\{ (1-x+y_l-y_u) \left[6 \alpha_1 x (1-x+y_l)^2 \right. \right. \nonumber \\ &+& 
 12 \alpha_2 (2-x)(1-x+y_l)\sqrt{y_l y_u}  +
\alpha_3 (2 x^3-x^2(5+5 y_l+y_u)-4 y_l(1+y_l+2 y_u) \nonumber \\ &+&
\left.  x(3+10y_l+3y_l^2+3y_u+3y_l y_u)) \right] 
+ \left. 24 \alpha_4 x (1-x+y_l)^2 \sqrt{y_l y_u} \right\}
\end{eqnarray} 
with $~2\sqrt{y_y}<x<1+y_l-y_u , ~ y_l=\left(\frac{m_l}{m_N}\right)^2, ~y_u=\left(\frac{m_u}{m_N}\right)^2$
and the coefficients $\alpha_{1,..,4}$ take the expressions:
\begin{eqnarray*}
\alpha_1&=&\left(\alpha_{s_1,i_u}^2+\alpha_{s_2,i_u}^2-
\alpha_{s_2,i_u}\alpha_{s_3,i_u} \right)\delta_{i_u,i_l}
\nonumber \\
\alpha_2&=&\left(\alpha_{s_1,i_u} \alpha_{W,i_l}\frac{y_W(1-x+y_l-y_W)}{(1-x+y_l-y_W)^2+y_W y_{\Gamma_W}}-
\alpha_{s_3,i_u}\alpha_{V_{0},i_u}\right)\delta_{i_u,i_l}
\nonumber \\
\alpha_3&=&\left(\alpha_{s_3,i_u}^2+4 \alpha_{V_{0},i_u}^2\right)\delta_{i_u, i_l}+
4 \alpha_{W,i_l}^2\frac{y_W^2(1-x+y_l-y_W)}{(1-x+y_l-y_W)^2+y_W y_{\Gamma_W}}
\nonumber \\
\alpha_4&=& \alpha_{s_2,i_u} \alpha_{v,i_u} \delta_{i_u,i_l}
\end{eqnarray*}
with $y_W=\left(\frac{m_W}{m_N}\right)^2,~ y_{\Gamma_W}=\left(\frac{\Gamma_W}{m_N}\right)^2$.
\begin{eqnarray}\label{dquarks_2}
\frac{d\Gamma}{dx}^{(N \rightarrow \nu d d)}&=&\frac{m_N}{128\pi^3}
\left( \frac{m_N}{\Lambda}\right)^4
\frac{x^2}{4}\frac{\sqrt{(1-x)(1-x-4y_d)}}{(1-x)^2}
\left\{\alpha_{s_3,i_d}^2 (3+x(-5+2 x+2 y_d))  \right. \nonumber \\  &+&  \left. 
6\left(\alpha_{s_2,i_d}^2-\alpha_{s_2,i_d}\alpha_{s_3,i_d}\right)
(1-x)(1-x-2 y_d) 
 \right\} \delta_{i_d,i_l}
\end{eqnarray}
with $0<x<1-4 y_d , ~y_d=\left(\frac{m_d}{m_N}\right)^2$
\begin{eqnarray}\label{dquarks_3}
\frac{d\Gamma}{dx}^{(N \rightarrow \nu u
u)}&=&\frac{m_N}{128\pi^3}\left( \frac{m_N}{\Lambda}\right)^4  \alpha_{s_1,i_u}^2
 \frac32 x^2 \sqrt{1-\frac{4 y_u}{(1-x)}}(1-x-2 y_u) \delta_{i_u,i_l} 
\end{eqnarray}
with $0<x<1-4 y_u$.

And the purely leptonic decay:
\begin{eqnarray}\label{dleptons}
\frac{d\Gamma}{dx}^{(N\rightarrow l^{+} leptons)}&=&\frac{m_N }{1536\pi^3}
\left( \frac{m_N}{\Lambda} \right)^4 \frac{(1-x+y_l-y_{l^{\prime}})^2}{(1-x+y_l)^3} 
 x \left[ \alpha_1 P(x)-\alpha_2 R(x) \right] 
\end{eqnarray}
with $2\sqrt{y_l} < x < 1+y_l-y_{l^{\prime}} ,~ y_l=\left(\frac{m_l}{m_N}\right)^2 , ~ y_{l^{\prime}}=\left(\frac{m_l^{\prime}}{m_N}\right)^2$
and $\alpha_{1,2}$ and the terms $P(x)$, $R(x)$ take the expressions:
\begin{eqnarray*}
\alpha_1&=&\alpha_{s_0,i_{l^{\prime}}}^2 \delta_{i_l,i_{l^{\prime}}}+
\frac{4 \alpha_W^2 y_W^2}{(1-x+y_l-y_W)^2+y_W y_{\Gamma_W}}
\nonumber \\
\alpha_2&=&12 \alpha_{s_0,i_{l^{\prime}}}\alpha_{W,i_l} 
\frac{(1-x+y_l-y_W}{(1-x+y_l-y_W)^2+y_W y_{\Gamma_W}} \delta_{i_l,i_{l^{\prime}}}
\nonumber \\
P(x)&=&2x^3-x^2(5+5 y_l+y_{l^{\prime}})-4 y_l (1+y_l+2y_{l^{\prime}})+
x(3+10 y_l + 3 y_l^2+3y_{l^{\prime}}+3 y_l y_{l^{\prime}})
\nonumber \\
R(x) &=& (2-x)(1-x+y_l) \sqrt{y_l y_{l^{\prime}}}.
\end{eqnarray*}
In the last expressions $x= 2 p^0_{lepton}/m_N$. 

Finally, in the considered mass range, the one-loop operators in the Lagrangian (\ref{leff_1loop}), induce the decay of $N$ to neutrino and photon:
\begin{eqnarray}
\label{photon_neutrino}
\Gamma^{N\rightarrow \nu_i (\bar \nu_{i}) A}=\frac1{2\pi}\left(\frac{v^2}{m_N}\right)\left(\frac{m_N}{\Lambda}\right)^4(\alpha_{L_1}^{(i)}c_W+\alpha_{L_3}^{(i)}s_W)^2
\end{eqnarray}

This decay mode leads to an interesting phenomenology, as will be shown in the following sections.

\subsection{Bounds on the couplings $\alpha^i_{\mathcal{O}}$}\label{Bounds}

Existent bounds on right-handed heavy Majorana neutrinos (often called ``sterile'', as they are $SU(2)$ singlets) are generally imposed on the parameters representing the mixing between them and the light left-handed ordinary neutrinos (``active''). Very recent reviews \cite{Deppisch:2015qwa, Antusch:2015mia,  Drewes:2015iva} summarize in general phenomenological approaches the existing experimental bounds, considering low scale minimal seesaw models, parameterized by a single heavy neutrino mass scale $M_{N}$ and a light-heavy mixing $U_{lN}$, where $l$ indicates the lepton flavor. The mentioned mixings are constrained experimentally by neutrinoless double beta decay, electroweak precision tests, low energy observables as rare lepton number violating (LNV) decays of mesons, peak searches in meson decays and beam dump experiments, as well as direct collider searches involving Z decays. Also, previous analysis \cite{delAguila:2006dx, delAguila:2008iz} refer in general to similar heavy neutrino-standard boson interaction structures, e.g.:
\begin{eqnarray}
\label{lw}
 \mathcal L_W = -\frac{g}{\sqrt{2}}  \overline l 
\gamma^{\mu} U_{lN} P_L N W_{\mu} + h.c.
\end{eqnarray}
\begin{eqnarray}
\label{lz}
 \mathcal L_Z = -\frac{g}{2 c_{W}} \overline \nu_{L} 
\gamma^{\mu} U_{lN} P_L N Z_{\mu} + h.c.
\end{eqnarray}
The effects of this modification on the weak currents are studied, as they lead to corresponding variations in the weak bosons decay rates and $W$ and $Z$ mediated processes involved in the existing experimental tests, specially in colliders \cite{ Antusch:2015mia, Antusch:2014woa, delAguila:2005pf, Bray:2005wv, Langacker:1988up, Nardi:1994iv, Decamp:1991uy, Abreu:1996pa}.

In the effective Lagrangian framework we are studying, the heavy Majorana neutrino couples to the three fermion family flavors with couplings dependent on the new ultraviolet physics scale $\Lambda$ and the constants $\alpha^{(i)}_{\mathcal O}$, where $i$ labels the families and $\mathcal{O}$ the operators. 

The operators presented in \eqref{eq:ope1} lead to a term in the effective Lagrangian \eqref{leff_svb} that can be compared to the interaction in \eqref{lw}, and a relation between the coupling $\alpha^{(i)}_{W}$ and the mixing $U_{lN}$ was derived in \cite{delAguila:2008ir}: $ U_{l_{i}N}\simeq \frac{\alpha^{(i)}_{W}v^2}{2\Lambda^2}$, while no operators lead to a term that can be directly related -with the same Lorentz-Dirac structure- to the interaction in \eqref{lz} (nor at tree or one-loop level). Some terms in the Lagrangian \eqref{leff_1loop} contribute to the $ZN\nu$ coupling, but as they are generated at one-loop level in the ultraviolet underlying theory, they are suppressed by a $1/16 \pi^2$ factor. 

In consequence, we take a conservative approach. In order to keep the analysis as simple as possible, but with the aim to put reliable bounds on our effective couplings, in this work we relate the mixing angle between light and heavy neutrinos ($U_{eN}$, $U_{\mu N}$, $U_{\tau N}$) with the couplings as $U \simeq \frac{\alpha^{(i)}_{\mathcal O} v^2}{2\Lambda^2}$ where $v$ corresponds to the vacuum expectation value: $v=250$ G$e$V. As we will explain shortly, we consider two situations in which the different bounds applies to the couplings.

Some of the considered operators contribute directly to the neutrinoless double beta decay ($0\nu\beta\beta$-decay) and thus the corresponding coupling constants, involving the first fermion family $i=1$, are restricted by strong bounds. We explicitly calculated the implications for the effective couplings in our Lagrangian.

In a general way, the following effective interaction Hamiltonian can be considered:
\begin{equation}
\label{heff}
\mathcal{H}=G_{eff} \; \bar u \Gamma d \;\; \bar e
\Gamma N + h.c.
\end{equation}
where $\Gamma$ represents a general Lorentz-Dirac structure.
Following the development presented in \cite{Mohapatra:1998ye} and using the
most stringent limit on the lifetime for $0\nu\beta\beta$-decay $\tau_{{0\nu}_{\beta\beta}} \geq 2.1 \times 10^{25}$ years
obtained by the Gerda Collaboration \cite{Macolino:2013ifa} we
have obtained the following bounds on the $G_{eff}$
\begin{equation}
G_{eff} \leq 7.8 \times 10^{-8} \left( \frac{m_N}{100 GeV}
\right)^{1/2} GeV^{-2}.
\end{equation}

The lowest order contribution to $0\nu_{\beta \beta}$-decay from the considered effective operators comes from those containing the $W$ field and the 4-fermion operators with quarks $u$, $d$, the lepton $e$ and the Majorana neutrino $N$. These operators contribute to the effective Hamiltonian \eqref{heff}, with
$G_{eff}=\frac{\alpha}{\Lambda^2}$. Thus we can translate the limit coming from  $G_{eff}$ on $\alpha^{(1)}_{\mathcal{O}}$ which, for $\Lambda=1$ T$e$V, is
\begin{equation}
\alpha^{bound}_{0\nu\beta\beta} \leq 7.993 \times 10^{-2}
\left(\frac{m_N}{100 ~GeV}\right)^{1/2}.
\end{equation}

\begin{figure*}[h]
\centering
\subfloat[]{\label{fig:branching}\includegraphics[totalheight=5.8cm]{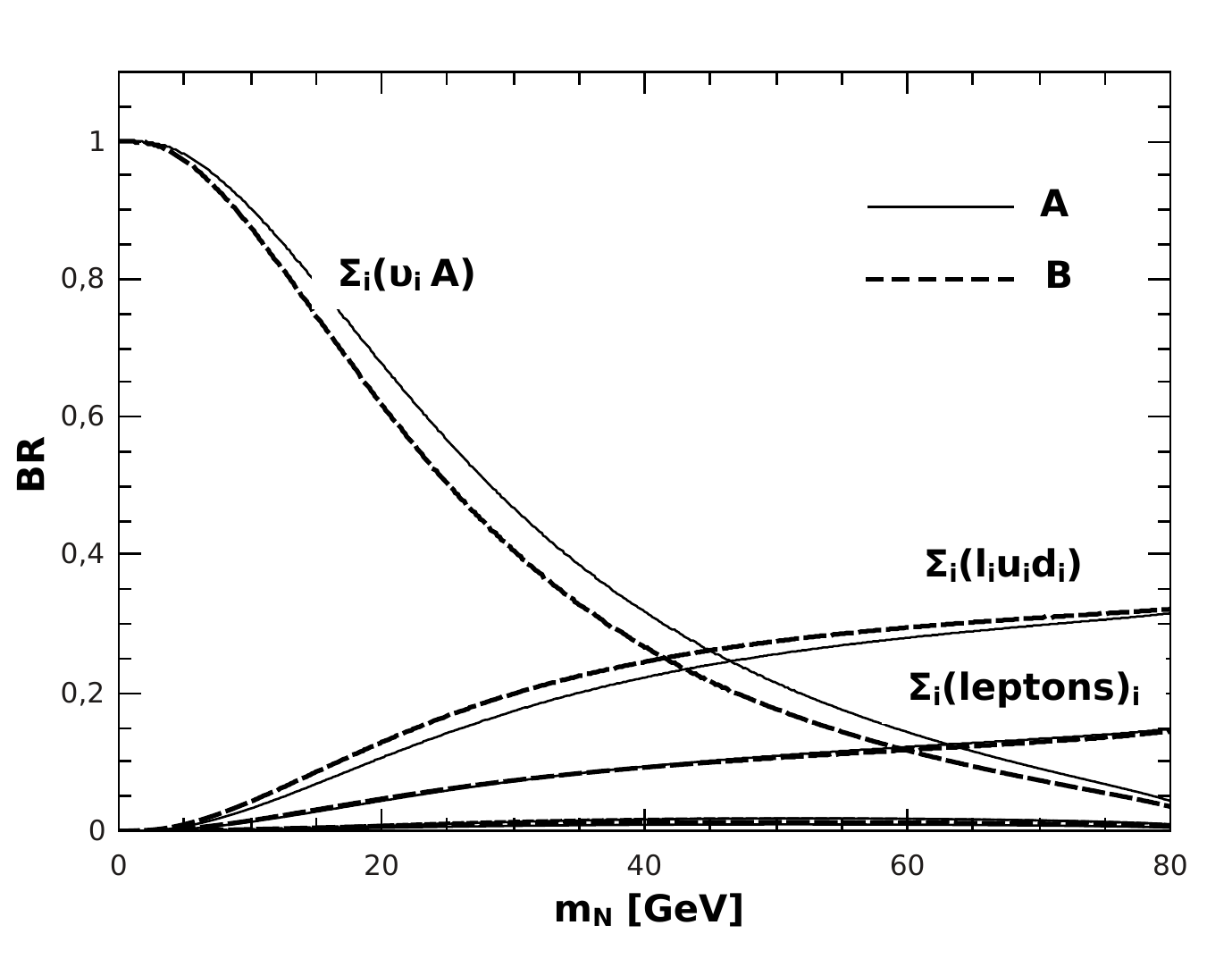}}~
\subfloat[]{\label{fig:width}\includegraphics[totalheight=5.8cm]{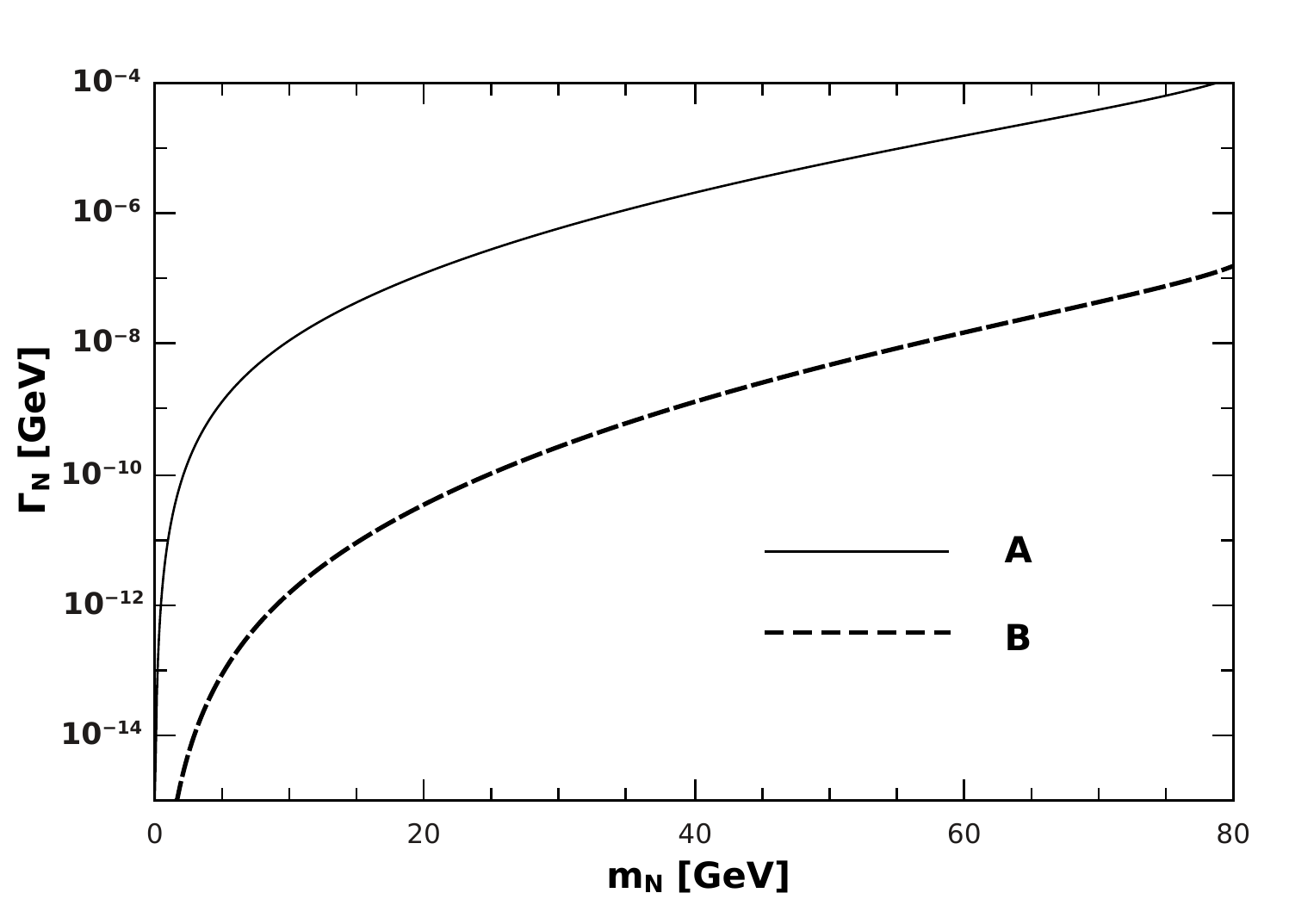}}
\caption{ (a)The Branching ratios for the Majorana neutrino decay with coupling constants in the set {\bf A} (solid lines) and set {\bf B} (dashed lines). The unlabeled curves correspond to the decay $N \rightarrow \nu uu$ and $N \rightarrow \nu dd$. (b) Total width for the two combinations of coupling constants defined in the text (set {\bf A} and set {\bf B}) and $\Lambda = 1\,$ T$e$V.}\label{fig:Br_Wth}
\end{figure*}

As will be explained in Sec \ref{applications}, the relevant Majorana neutrino mass range for considering this heavy neutral particle as a solution to the MiniBooNE anomaly is $400 MeV < m_N < 600 MeV$ \cite{Gninenko:2009ks}. The experimental bounds for this mass values are exhaustively discussed in \cite{Deppisch:2015qwa} and references therein. Taking into account that the MiniBooNE experiment deals with muon-type neutrinos, we now discuss the bounds on the $U_{\mu N}$ mixings, which are not constrained by $0\nu_{\beta \beta}$-decay, and are most restrictive than the existing ones for the third fermion family.

As can be seen in \cite{Deppisch:2015qwa}, the existing bounds for $U_{\mu N}$ for $m_{N}\simeq 500$ M$e$V come from beam dump experiments as NuTeV \cite{Vaitaitis:1999wq}, CHARM II \cite{Vilain:1994vg} and BEBC \cite{CooperSarkar:1985nh}, rare lepton number violating (LNV) meson decays at LHCb \cite{Aaij:2014aba} and from colliders as those from DELPHI \cite{Abreu:1996pa}. 
In the case of the heavy Majorana neutrino with effective interactions we are considering, the clear dominance of the neutrino plus photon channel found in \eqref{photon_neutrino} makes the beam dump and rare LNV experiments bounds inapplicable, as this decay mode to invisible particles is not considered in those analysis, and can considerably alter the number of events found for $N$ decays inside the detectors \cite{Deppisch:2015qwa, Drewes:2015iva}.  

In the light of this discussion, we consider the bounds from DELPHI \cite{Abreu:1996pa}, following the treatment made in \cite{delAguila:2006dx}. In our case, for only one heavy Majorana neutrino we have: $\Omega_{l l'} = U_{l N} U_{l' N}$ and the allowed values for the mixings are of order:
\begin{eqnarray}\label{collbounds}
U^2_{\mu N} \lesssim 5\times 10^{-3}
\end{eqnarray}
For the Lepton-Flavor-Violating processes e.g. $\mu \rightarrow e \gamma$, $\mu \rightarrow e e e$ and $\tau \rightarrow e e e$, which are induced by the quantum effect of the heavy neutrinos, we have very weak bounds for $m_N < m_W$ \cite{Antusch:2015mia, Drewes:2015iva, Tommasini:1995ii}.

Thus, the bound in \eqref{collbounds} can be translated to the constants $\alpha$, and we have for $\Lambda=1$ T$e$V 
\begin{eqnarray}\label{collbound_value}
\alpha^{bound}_{Coll} \leq 2.3
\end{eqnarray}

For completion we have explicitly calculated the bounds that can be inferred from the single $Z\rightarrow \nu N$ and pair $Z\rightarrow N~N$ ``excited'' neutrino production searches at LEP \cite{Decamp:1991uy}. The first process can be generated by one-loop level effective operators \eqref{eq:ope3} giving the terms in the Lagrangian \eqref{leff_1loop}. As the one loop level couplings are supressed by the factor $1/16 \pi^2$, the corresponding bound for the couplings $(\alpha^{tree} v^2/2\Lambda^2)^2$ is absorbed by the $(16 \pi^2)^2$ multpliying the bounds, so that the collider \eqref{collbound_value} value is still more stringent. 
It is important to mention that other effective operators (4-fermion operators in \eqref{leff_4-f}) contribute to the $\nu N$ and $N N$ production at LEP, but at the $Z$ peak they give less restrictive bounds than the ones in \eqref{collbounds}. 
For the decay $Z\rightarrow N~N$, we have a direct contribution from the tree level operator $\mathcal{O}_{NN\phi}$, giving 
\begin{eqnarray}
 \Gamma(Z\rightarrow N N)= \frac{\alpha^2_Z c^2_{W}}{96 \pi s^2_{W}} \left(\frac{v}{\Lambda}\right)^4 m_Z.
\end{eqnarray}
A conservative limit for any $m_N$ mass is $Br(Z\rightarrow NN)Br^{2}(N\rightarrow \nu (\bar \nu) \gamma)<5\times 10^{-5}$ \cite{Decamp:1991uy}. 
This result is model-independent and holds for the production of a pair of heavy neutral objects decaying into a photon and a light invisible particle.
For the low $m_{N}$ values considered in this work, we can take $Br(N\rightarrow \nu (\bar \nu) \gamma)\simeq 1$ and the corresponding bound is $(\frac{\alpha_{Z} v^2}{2 \Lambda^2})^2<3.0 \times 10^{-5}$, more restricting than the bound in \eqref{collbounds}, but not taken into account, as the corresponding operator does not contribute to the $N$ decay.

In order to simplify the discussion, for the numerical evaluation we only consider the two following situations. In the set we call {\bf A} the couplings associated to the operators that contribute to the $0\nu\beta\beta$-decay ($\mathcal{O}_{N e \phi}$, $\mathcal{O}_{d u N e}$, $\mathcal{O}_{Q u N L}$, $\mathcal{O}_{L N Q d}$ and  $\mathcal{O}_{Q N L d}$) for the fisrst family are restricted to the corresponding bound $\alpha^{bound}_{0\nu\beta\beta}$ and the other constants are restricted to the bound determined by colliders  $\alpha^{bound}_{Coll}$. In the case of the set called {\bf B} all the couplings are restricted to the $0\nu\beta\beta$ bound $\alpha^{bound}_{0\nu\beta\beta}$ which is the most stringent. For the  1-loop generated operators we consider the coupling constant as $1/(16 \pi^2)$ times the corresponding tree level coupling: $\alpha^{1-loop}=\alpha^{tree}/(16 \pi^2)$. Thus, for the operators $\mathcal{O}_{DN}$, $\mathcal{O}_{NW}$ and $\mathcal{O}_{\bar DN}$, which contribute to $0\nu\beta\beta$ we have 
\begin{equation}
\alpha^{(1)}_{L_2}, \alpha^{(1)}_{L_3}, \alpha^{(1)}_{L_4} \sim \frac{1}{16\pi^2} \alpha^{bound}_{0\nu\beta\beta}
\end{equation}
for fermions of the first family. For the remaining operators we take
\begin{equation}
\alpha \sim \alpha^{bound}_{Coll} , \alpha^{bound}_{0\nu\beta\beta} 
\end{equation}
in the sets {\bf A} and {\bf B} respectively.

In Fig. \ref{fig:Br_Wth} we show the results for the Majorana neutrino decay presented in Sec. \ref{Decay_Widths}.
Figure \ref{fig:branching} shows the branching ratio as a function of the Majorana neutrino mass $m_N$. The decay is calculated for different values of the constants $\alpha^i_{\mathcal O}$. We show the branching ratios for both sets {\bf A} and {\bf B}. It can be seen that, for low masses, the dominant channel is the decay of $N$ to photon and neutrino. Figure \ref{fig:width} shows the total decay width dependence on the mass for both coupling sets considered.  

Taking the values of the couplings $\alpha^{(i)}$ to be equal for every family $i$, and also for every tree level coupling $\alpha^{tree}$, and taking the one-loop generated couplings as $\alpha^{1-loop}=\alpha^{tree}/16\pi^2$, we derived an approximated expression for the ratio between the widths $\Gamma (N \rightarrow \nu(\bar \nu) A)$ in \eqref{photon_neutrino} and $\Gamma (N\rightarrow l^{+} \bar u d)$ in \eqref{dquarks_1}:
\begin{eqnarray}
 \frac{\Gamma^{(N \rightarrow \nu(\bar \nu) A)}}{\Gamma^{(N\rightarrow l^{+} \bar u d)}}\rightarrow \frac{2}{15 \pi} \left(\frac{v}{m_{N}}\right)^2 \left( c_{W}+s_{W}\right)^2
\end{eqnarray}
This limiting value explains the behavior found in Fig.\ref{fig:Br_Wth} for low Majorana neutrino masses, showing the neutrino plus photon decay channel is clearly dominating. This is an interesting fact since we have a new source of photons, leading to a very rich phenomenology discussed in the next section.


\section{Application to neutrino-related questions}\label{applications}

Searches for heavy neutrinos often rely on their possibility to decay to detectable particles. The interpretation of the corresponding results for such searches requires a model for the decay of the heavy neutrino. Several explanations to different kind of problems seem related to weakly interacting neutral particles, like new neutrinos. In particular the MiniBooNE \cite{AguilarArevalo:2007it} anomaly or the observation of sub-horizontal air-showers by Cherenkov telescope SHALON \cite{Sinitsyna:2013hmn} have possible explanations by long lived neutral particles like the one studied in this work.

The MiniBooNE experiment was built to search for $\nu_{\mu} \rightarrow \nu_{e}$ conversion, in order to confirm or refute the previous results of LNSD, which were inconsistent with global neutrino oscillation data \cite{Aguilar:2001ty}. The MiniBooNE anomaly consists in an unexplained excess of low energy electron-like events in charge-current quasi-elastic electron neutrino events over the expected standard neutrino interactions \cite{AguilarArevalo:2007it, AguilarArevalo:2008rc}. 

This excess of electron-like events could be caused by the decay of a heavy neutrino. This solution was proposed by Gninenko \cite{Gninenko:2009ks} in a model with sterile neutrino mixed with the standard neutrinos by a matrix $U$. He finds that $N$ with
\begin{eqnarray} \label{miniboone_Gninenko}
&& 400 MeV < m_N < 600 MeV \nonumber \\
&& 10^{-3} < \vert U_{\mu N} \vert^2 < 4 \, 10^{-3} \nonumber \\
&& 10^{-11} s < \tau_N < 10^{-9} s
\end{eqnarray}
could explain the anomaly, as the excess of electron-like events in the $\nu_{\mu}$ beam could be caused by the decay of a heavy neutrino with a radiative dominant decay mode $N \rightarrow \nu \gamma$ where the final photon would be converted into an $e^+ e^-$ pair with a small opening angle, indistinguishable from an electron in the detector. This is called a converted photon. 

The Gninenko analysis is based on the assumption that the heavy neutrino radiative decay is dominant. The effect of the mentioned strong radiative decay is the flux attenuation by $N$ decay and then the decrease of the signal events in the detector. The consequences are less restrictive bounds on $\vert U_{\mu N} \vert^2$ \cite{Deppisch:2015qwa, Drewes:2015iva}, as we explained in sec.\ref{Bounds}. The proposal is then that the excess of events observed by MiniBooNE could originate from converted photons and not from electrons. The future experiment MicroBooNE will provide a test to this proposal, as it will be able to separate photons from electrons or positrons \cite{Chen:2007ae}. 

In the context of the effective interactions considered in this work, one has to check if the $N\rightarrow \nu A$ is the dominant decay, by comparing the decay of $N$ to pions, which is the correct hadronic final state for the low masses studied here. We have found that the corresponding decay is mainly given by
\begin{eqnarray}
\Gamma^{N\rightarrow l_i^+ \pi^-}&=&\frac{G_F^2}{8\pi}\left(\frac{\alpha_W^{(i)} v^2}{2 \Lambda^2} \right)^2 f_{\pi}^2 m_N^3
\left[(1-\frac{m_l^2}{m_N^2})^2-\frac{m_{\pi}^2}{m_N^2}\left(1+\frac{m_l^2}{m_N^2} \right) \right] \times
\nonumber \\
&&\sqrt{\left(1+\frac{m_l^2}{m_N^2}-\frac{m_{\pi}^2}{m_N^2}\right)^2-4 \frac{m_l^2}{m_N^2}}.
\end{eqnarray}

In the mass range proposed \cite{Gninenko:2009ks} we find that the ratio of the branching ratios for the different decay channels is $Br(N\rightarrow l_i^+ \pi)/Br(N\rightarrow \nu (\bar\nu) A) \simeq 8 \times 10^{-6}$ and  $Br(N\rightarrow leptons)/Br(N\rightarrow \nu (\bar\nu) A)\simeq 4\times 10^{-6}$ thus confirming the dominance of the radiative decay $N\rightarrow \nu A$. 

\begin{figure*}[htbp]
\begin{center}
\includegraphics[totalheight=5cm]{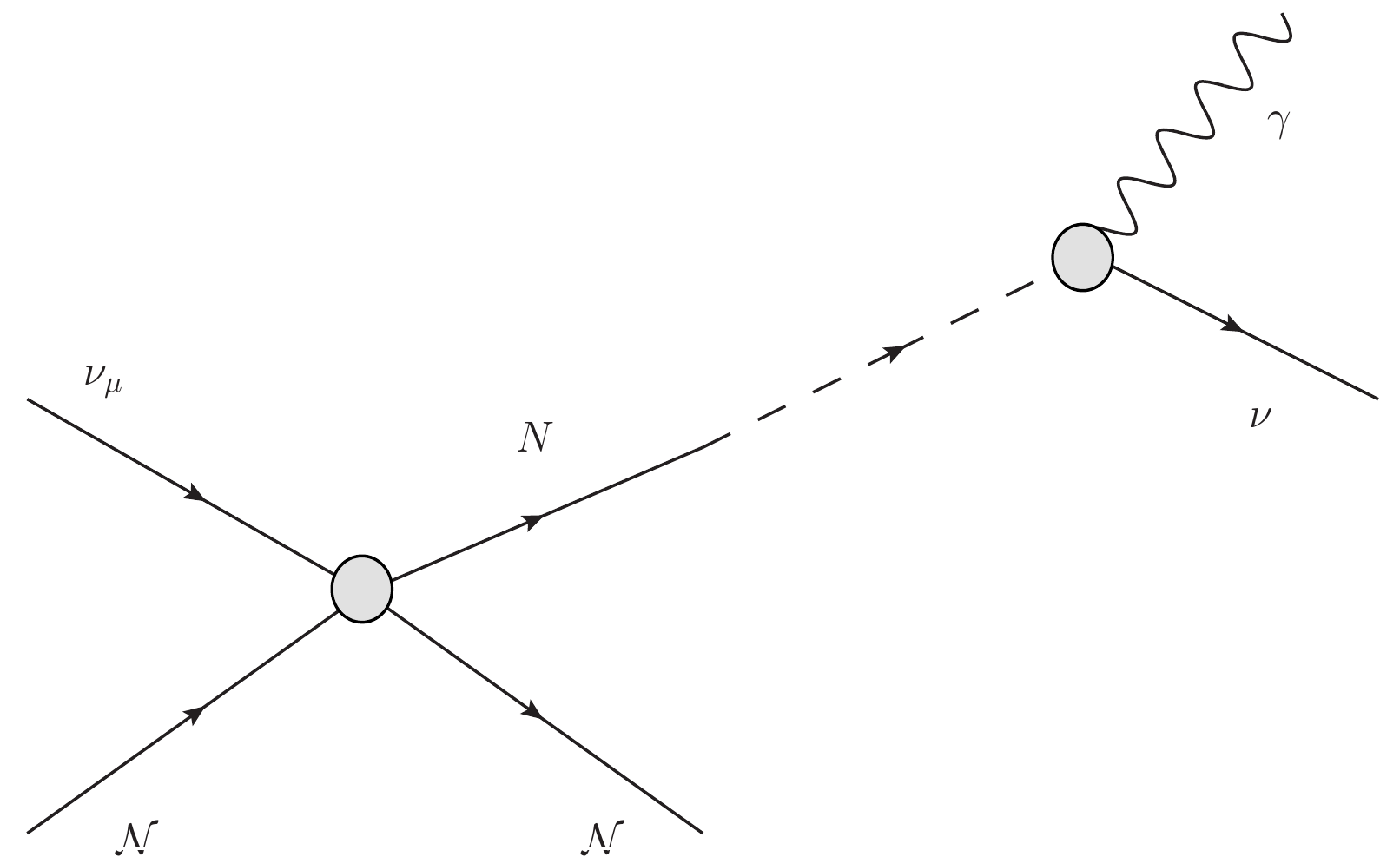}
\caption{\label{fig:NC} Production process of extra neutrino $N$ by effective interactions and their subsequent decay.}
\end{center}
\end{figure*}

The heavy neutrino $N$ could be directly produced by the $\nu_{\mu}$ in neutrino-nucleon reactions by the effective operators $\mathcal{O}_{Q u NL}$, $\mathcal{O}_{LNQ d}$ and $\mathcal{O}_{QNL d}$, with the subsequent decay and photon conversion as we show in Fig.(\ref{fig:NC}).

The excess of $\nu_e$ events is related to the relative magnitude between the Standard Model Neutral Current (SM NC) $\nu_{\mu}\mathcal{N}\rightarrow \nu_{\mu}\mathcal{N}$ process and the effective NC-like $N$ production $\nu_{\mu}\mathcal{N}\rightarrow N \mathcal{N}$ being $\mathcal{N}$ a nucleon. For the effective operator we have a 4-fermion contribution with intensity $\alpha/\Lambda^2$ and for the SM-NC $g^2/(4 m_W^2)$.
Then the amplitude ratio is $\mathcal{K}=\alpha v^2/2\Lambda^2$ and the $N$ production is weighed by the factor $\mathcal{K}^2=(\alpha v^2/2\Lambda^2)^2$ relative to the SM-NC $\nu_{\mu}$ scattering. The constant $\mathcal{K}^2$ plays the role of the mixing matrix $U_{\mu N}^2$ in the Gninenko \cite{Gninenko:2009ks} work, and then the value $U_{\mu N}^2$ found in \eqref{miniboone_Gninenko} is consistent with the allowed value by the collider bound of \eqref{collbounds} \cite{Bergmann:1998rg}.

\begin{figure*}[htbp]
\centering
\includegraphics[totalheight=8cm]{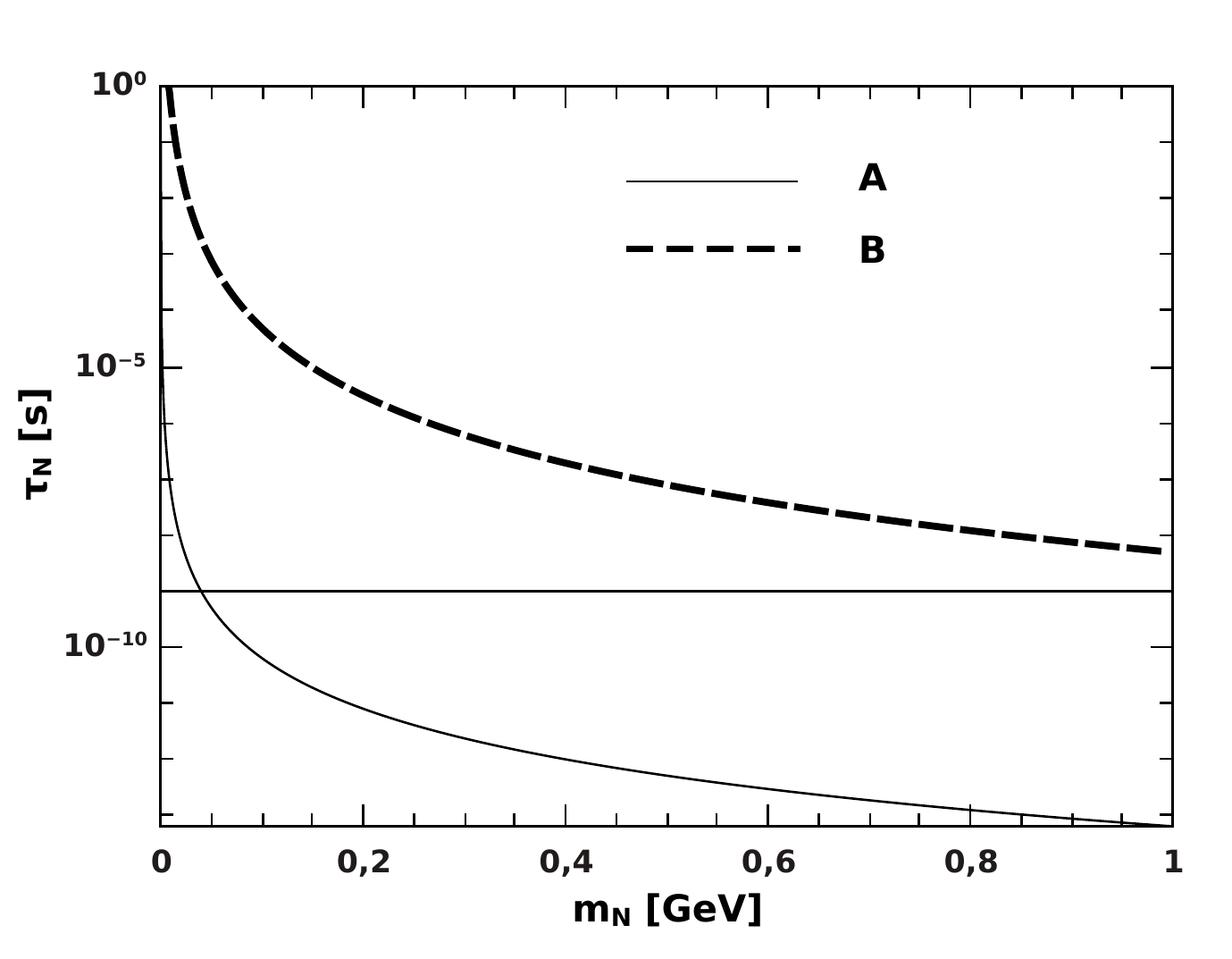}
\caption{\label{fig:tau} $N$ lifetime as a function of its mass for the sets of coupling constants labeled {\bf A} and {\bf B} for $\Lambda = 1$ T$e$V. For $\Lambda > 1$ T$e$V the values allowed for the lifetime correspond to the region upwards the curves. The horizontal solid line corresponds to the limit value found in the Gninenko solution for the MiniBooNE anomaly ($\tau_N<10^{-9} s$).}
\end{figure*}

The constraint for the lifetime of the heavy neutrino in \eqref{miniboone_Gninenko} must also be fulfilled in order to consider the $N$ effective radiative decay as an alternative explanation for the MiniBooNE anomaly. In Fig.(\ref{fig:tau}) we show the lifetime $\tau_N$ as a function of $m_N$ for the sets {\bf A} and {\bf B} and for $\Lambda=1$ T$e$V. In the case of $\Lambda > 1$ T$e$V the allowed region is upwards the curves. Thus, we can see a region compatible with set {\bf A} where $\tau_N < 10^{-9}\,s$ as in the solution proposed by Gninenko.

\begin{figure*}[h]
\centering
\subfloat[$N$ Decay Length]{\label{fig:ldeclm}\includegraphics[totalheight=5.8cm]{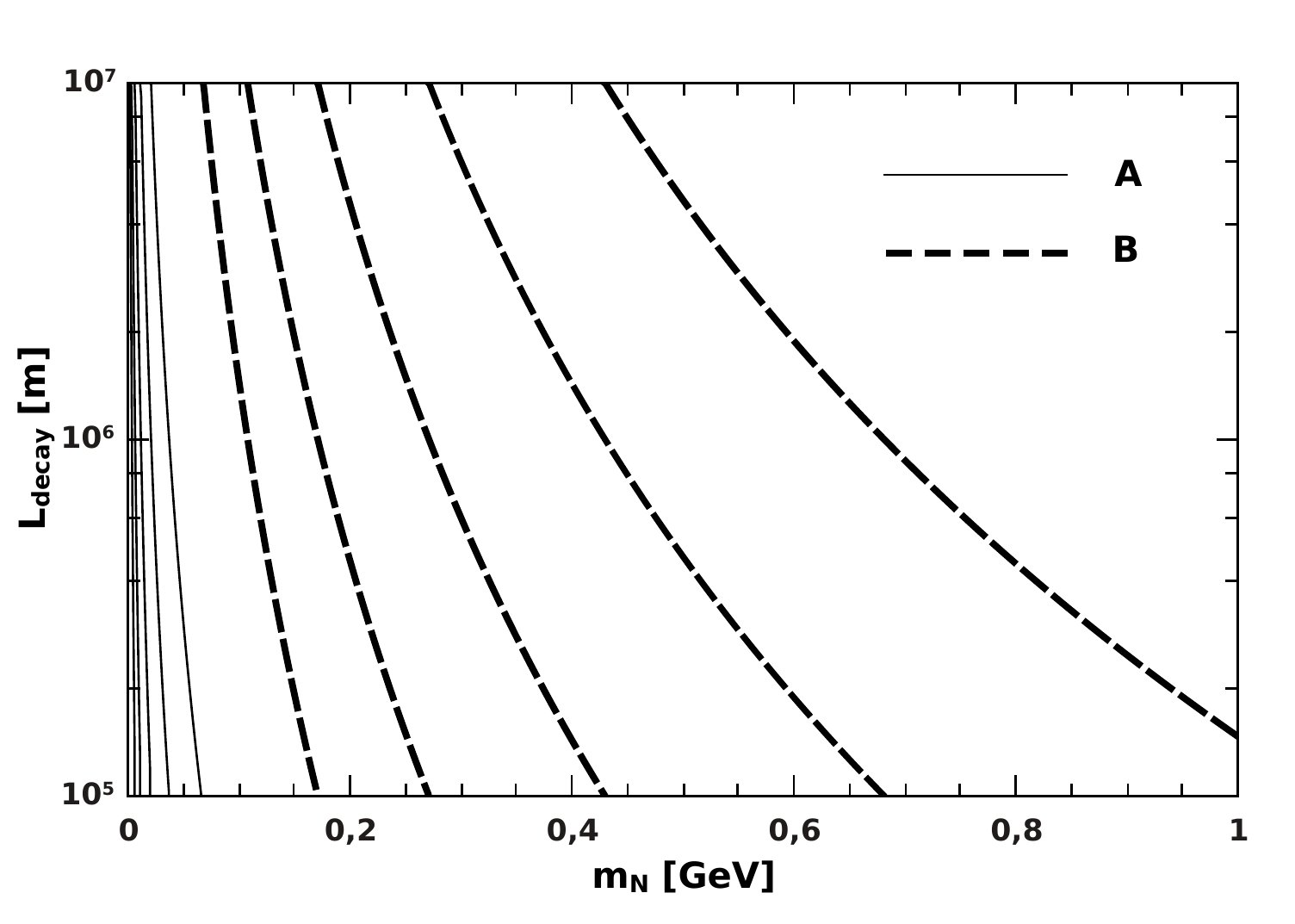}}~
\subfloat[$N$ Decay Length (higher masses)]{\label{fig:ldecvd}\includegraphics[totalheight=5.8cm]{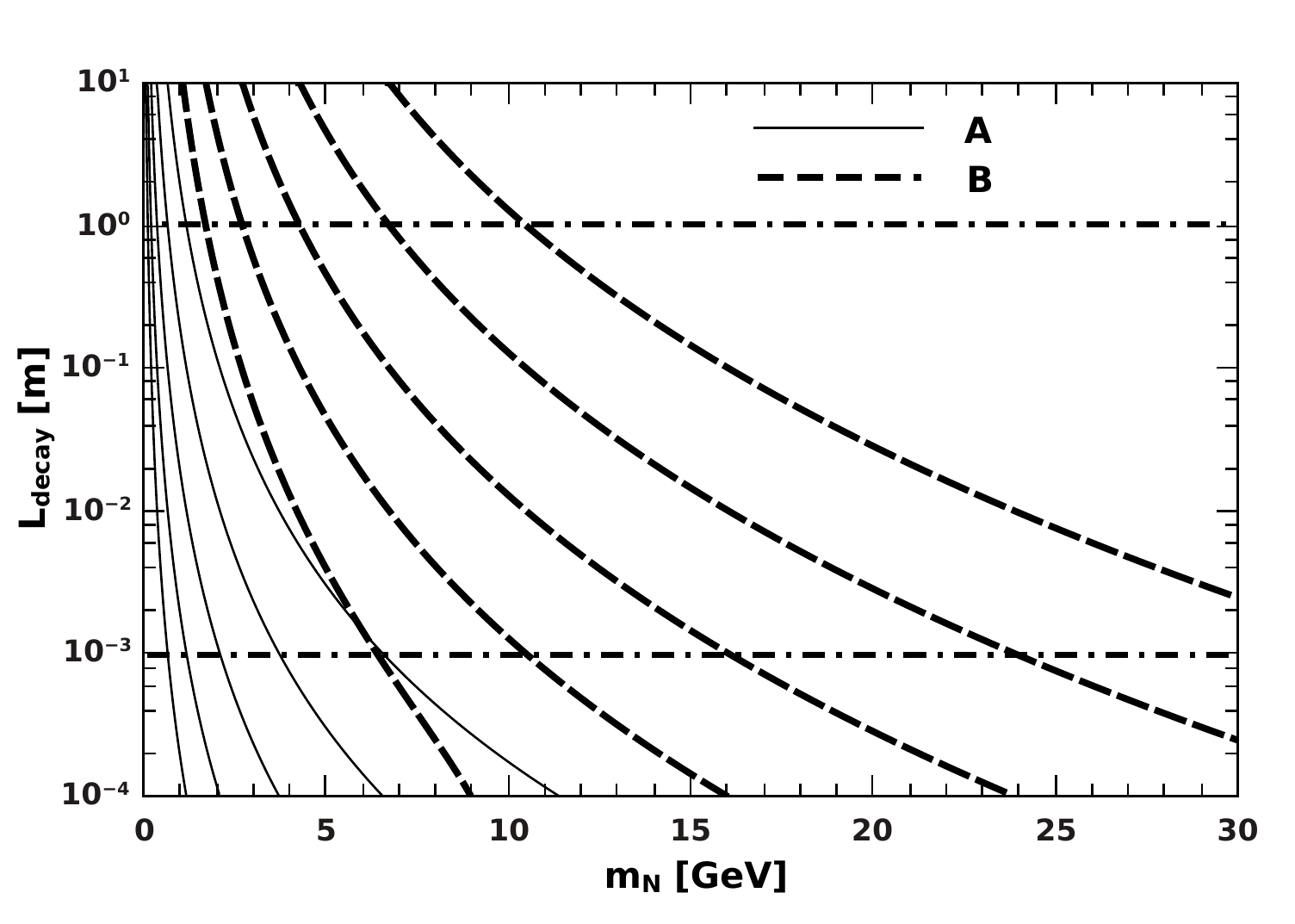}}
\caption{Decay length for different neutrino energies as a function of the neutrino mass for the coupling constant sets {\bf A} (solid lines) and {\bf B} (dashed lines) for $\Lambda=1$ T$e$V. The energies, $E=10^n$ G$e$V, vary from left to right with increasing $n$ (1-5). For $\Lambda > 1$ T$e$V the decay length corresponds to the top right region from the curves.}\label{fig:Dlenght}
\end{figure*}

As was previously mentioned, this kind of neutral particle which decays dominantly to neutrino and photon could be the explanation for several sub-horizontal events detected by the Cherenkov telescope SHALON as it was recently proposed in \cite{Sinitsyna:2013hmn}. In the cited work the authors propose that the solution could be a neutral and then penetrating long-lived massive particle able to cross $1000 ~km$ of rock and decay within the $7 ~km$ of air in front of the telescope. In Fig. \ref{fig:Dlenght} we show the decay length as a function of the heavy neutrino mass for different energies and couplings in the sets {\bf A} and {\bf B}. We can see that there is a region of the parameter space which could possibly explain the SHALON observations with a $l_{decay}\sim 1000 ~km$.

To conclude, a few words about the detectability of this particle in colliders like the LHC. 

Searches for neutral long-lived particles as the heavy neutrino proposed by \cite{Gninenko:2009ks} have been studied in the context of $\tau^{-}$ rare decays \cite{Dib:2011hc}, where the authors propose to search for events with two vertices, featuring the production and decay of the unstable neutrino $N$.
 The use of displaced vertices has also been proposed to search for sterile neutrinos at the LHC \cite{Helo:2013esa, Gago:2015vma}, for $N$ decaying to leptons and quarks or purely leptonically. Early displaced vertices searches are  reviewed in \cite{Graham:2012th}.

As we have shown, for the $N$ masses considered in this work the dominant decay is the radiative $N\rightarrow \nu \gamma$ channel, which can be observed by the signature of an isolated electromagnetic cluster together with missing transverse energy:
\begin{equation}
\gamma + E_T^{miss}
\end{equation}
where the photon originates in a displaced vertex.

New physics searches involving such final states have been performed at the LHC \cite{Aad:2014tda, Aad:2014gfa}, and it has been suggested that this signal could be enhanced with the combined use of missing transverse monentum plus photons and displaced vertices searching techniques \cite{Cui:2014twa, Biswas:2010yp}. The use of this technique will allow to probe parts of the parameter space which are inaccessible by other methods. The use of displaced vertices has the advantage that for decay lengths of the order of, very roughly $L \in (10^{-3}-1) ~m$, there is little standard model background. We find that decay lengths as the above mentioned for masses between $1-30$ G$e$V are possible in this model as we show in Fig.\ref{fig:Dlenght} for the sets {\bf A} and {\bf B}.

\section{Summary and conclusions}\label{conclusions}

We have calculated the decay widths and branching ratios for a relatively light heavy Majorana neutrino (with $m_{N}<m_{W}$) in an effective approach, considering its possible decays to fermions, quarks and photons, focusing on a relatively low neutrino mass range. We find that for masses below approximately $30$ G$e$V the dominant channel is the neutrino plus photon mode: $N \rightarrow \nu A$.  With this decay mode in mind, we explored the plausibility of considering it as an explanation for the MiniBooNE and SHALON anomalies. We checked that in the effective model the radiative decay is dominant respect to the lepton plus pion mode, and leads to values of the effective couplings $\alpha$ which are consistent with the mixing value $\vert U_{\mu N} \vert^2$ found by Gninenko \cite{Gninenko:2009ks} and with collider bounds \cite{Abreu:1996pa}. Also, we show that the Majorana neutrino lifetime also fits the limits in \cite{Gninenko:2009ks}. This kind of weakly interacting long-lived particle has also been proposed as an explanation for sub-horizontal events in the SHALON telescope \cite{Sinitsyna:2013hmn}, and we find that the $N$ decay length is compatible with the proposed explanation for part of our parameter space. This kind of particle could also be searched for in the  LHC, with the use of the displaced vertices technique, with little standard model background.\\

{\bf Acknowledgements}

We thank CONICET (Argentina) and Universidad Nacional de Mar del
Plata (Argentina); and PEDECIBA, ANII, and CSIC-UdelaR (Uruguay) for their 
financial supports.


\end{document}